\documentclass[%
 reprint,
 amsmath,amssymb,
 aps,
 pra,
 floatfix,
]{revtex4-1}

\usepackage{graphicx}
\usepackage{dcolumn}
\usepackage{bm}
\usepackage{braket}
\usepackage[inline]{enumitem}
\usepackage{float}

\begin{document}

\preprint{APS/123-QED}

\title{Measurement and feedback for cooling heavy levitated particles in low frequency traps
}

\author{L. S. Walker}
\author{G. R. M. Robb}
\author{A. J. Daley}
\affiliation{
 SUPA, Department of Physics, University of Strathclyde,
107 Rottenrow, Glasgow G4 0NG, United Kingdom
}

\date{\today}

\begin{abstract}
We consider a possible route to ground state cooling of a levitated nanoparticle, magnetically trapped by a strong permanent magnet, using a combination of measurement and feedback. The trap frequency of this system is much lower than those involving trapped ions or nano-mechanical resonators. Minimisation of environmental heating is therefore challenging as it requires control of the system on a timescale comparable to the inverse of the trap frequency. We show that these traps are an excellent platform for performing optimal feedback control via real-time state estimation, for the preparation of motional states with measurable quantum properties.
\end{abstract}

\maketitle


\section{\label{sec:Introduction}Introduction}

The ability to prepare and manipulate quantum states of nano-mechanical systems is of interest in metrology and for tests of fundamental quantum physics. Ground state cooling has already been achieved in cryogenic chambers with silicon membranes and other microwave devices\,\cite{Verhagen2012,Chan2011}. However, there is a desire to produce quantum states of motion with levitated particles that are not physically tethered to their surroundings, and which therefore have significantly longer decoherence times. If realised, these systems would be a platform for many novel experiments; tests of wave function collapse models \,\cite{Yin2013}, for ultra-sensitive metrology\,\cite{Geraci2010}, and to probe gravitational decoherence\,\cite{Albrecht2014}. Most of the progress towards preparing ground state systems has been made with optically levitated particles, where recent experiments are currently capable of detecting - and are limited by - photon shot noise\,\cite{Jain2016}.

Although optical traps are the most widely used for trapping microscopic particles, they can face problems with heating due to the high laser intensities\,\cite{Rahman2016} and the intrinsic noise associated with the trapping force. While optical traps are still very good for high-frequency scenarios, these difficulties become more severe at low frequencies. Magnetic traps are free from these problems and have recently been demonstrated as suitable for trapping and cooling nano-diamonds\,\cite{Hsu2016, OBrien2019}. The traps are typically three orders of magnitude larger than their optical counterparts, and consequently operate at much lower frequencies, of around $100$Hz as opposed to $100$kHz for an optical trap. This comes with the advantage of being able to hold and manipulate large particles, but also makes it unfeasible to cool on time-scales much longer than the relatively long oscillation period. Current experiments\,\cite{Jain2016,Hsu2016} have estimated the phonon reheating rate for these systems in high vacuum ($10^{-8}$ mbar) to be $\Gamma_{\text{th}}\approx100$Hz and it is expected that this will be significantly reduced at lower pressures. 

In this article we consider methods for improving the quantum measurement efficiency of levitated nano-particles, and go on to analyse how best to apply feedback and assess the fundamental cooling limits. Direct feedback of a position measurement in the form developed by Wiseman and Milburn\,\cite{Wiseman1994} has been shown to be effective in controlling the motion of optically levitated ions\,\cite{Bushev2006}, but we find it to be less suitable here. The cooling strategies employed with direct feedback rely heavily on a separation of time-scales between the damping and trap frequency that is impractical in larger traps. Instead, our starting point is to adapt the real time state estimation and the feedback strategies discussed by Doherty et al.\,\cite{A.C.Doherty1999} for use in this newly accessible low-frequency regime. 

Having considered several options for tracking a particle's position and momentum, we suggest making measurements in two steps. At first, scattered light from the particle can be imaged with a quadrant photo-diode, and an externally applied damping force can be used for cooling. After damping the particle's motion to sub-optical-wavelength amplitudes, significantly better resolution can be achieved by measuring how the particle scatters light into a mirror mode. An ideal candidate particle for future experiments would be an approximately spherical nano-diamond, which are of interest due to access to internal nitrogen-vacancies (NV). This second quantum handle on the particle is crucial for many proposed future experiments\,\cite{Yin2013,Albrecht2014} and may also provide a route to having fine control over micron, as opposed to nanometre, sized particles. We go on to show that the proposed methods could be used to produce motional states of microscopic oscillators with average phonon occupancy $\langle n\rangle < 3$ and state purity $P\approx0.44$, achievable with realistic measurement efficiencies for current experiments. This is a regime where it should already be possible to see signs of quantum behaviour in the particle motion, and could provide a starting point for preparation of more exotic macroscopic superposition states. With advances in isolation from environmental heating, and improvements in light collection efficiency, there are no fundamental limits to these techniques being used to reach the quantum motional ground state.

The rest of this article is organised as follows. In Sec. II we review the stochastic master equation that results from measuring the motion of a particle in front of a single mirror. In Sec. III we discuss the merits and limitations of various measurement schemes, and the practicality of real time state estimation. In Sec. IV we show the effectiveness of feedback by estimation, in cooling and squeezing mechanical motion. We conclude and present outlooks in Sec. V.

\section{\label{sec:model}Model}
Levitated, trapped particles for the purpose of cooling are, by design, simple oscillators. Our model describes the motion of a magnetically confined particle and its interaction with an optical probe beam. We will treat the internal dynamics of the light scattering process adiabatically, and model the particle as a point dipole in the Rayleigh regime - alternatives for larger particles will be discussed in the conclusions. The Hamiltonians of the freely oscillating particle, $H_{sys}$, the optical field, $H_F$, and the interaction Hamiltonian, $H_I$, are given by
\begin{equation}
H_{\text{sys}} = \frac{p^{2}}{2m} +\frac{m \omega^{2} x^{2}}{2} \text{,}
\end{equation}
\begin{equation}
H_{\text{F}} = \sum_{k}\hbar \omega_{k} b_{k}^{\dagger}b_{k},
\end{equation}
\begin{equation}
H_{\text{I}} = \sum_{k} \hbar \sqrt{\gamma}\big(b_{k}\exp(i \mathbf{k}.\mathbf{r})+b_{k}^{\dagger}\exp(-i \mathbf{k}.\mathbf{r})\big) \text{,}
\end{equation}
where $m$ is the particle mass, $\omega$ is the magnetic trap frequency, $\gamma$ is the scattering rate into each mode of the optical electric field, $b_k(b_k^{\dagger})$ is the usual quantised field mode amplitude, with wavenumber and angular frequency of k, $\omega_k$ respectively. The momentum recoil due to the scattered photons is represented by $\mathbf{k.r}$, where $\mathbf{r}$ is the particle's position. It is sufficient to model the motion of the particle in 1D, as although some cooling is often applied along each trap axis, the frequencies of each motional degree of freedom can be well separated and safely decoupled, as is done in current experiments \cite{Hsu2016}.

Continuous measurement theory allows for quantification of the disturbance caused to the particle in relation to the amount of position information carried away by the field\,\cite{Jacobs2006}. We will go on to discuss the merits and drawbacks of various measurement schemes, but first we outline the details of the method we assess to be the most suitable for magnetically levitated particles.

\subsection{\label{sec:detection}Motional side-band detection}
\begin{figure}[ht]
	\begin{center}
		\textbf{}\par\medskip
		\includegraphics[scale=0.4]{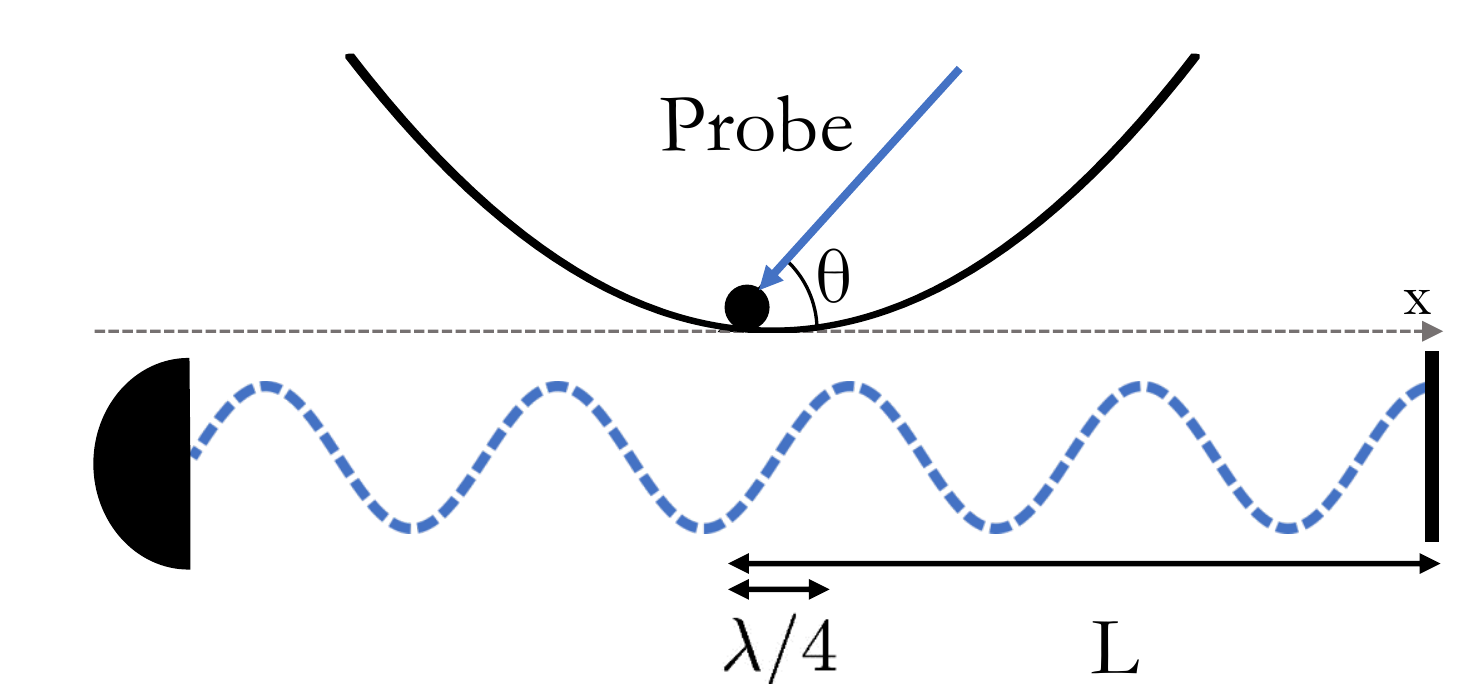}
		\caption{Sketch of apparatus for measuring the intensity of a standing wave, modulated by a particle's motion along its main trap axis $x$. The trap centre is marked a distance $L$ from the mirror, with the probe light incident at an angle $\theta$. The range of motion over which this measurement would be valid is restricted about a node of the standing light field, and has also been marked. With shaped illumination, light collection efficiencies $>15\%$ could be reasonably achieved.}
		\label{fig:setup}
	\end{center}
\end{figure}

The set-up we consider uses a mirror to introduce a standing wave mode across the levitated particle, where some of the scattered light from the illumination probe will be collected, as shown schematically in Fig.~\ref{fig:setup}. The mirrors here can be quite large, capturing a significant fraction of the light scattered along the primary trap axis. The particle motion adds side-bands to the spectrum of light scattered in the mirror mode, positioned at $\pm \omega$ from the optical frequency. Continuous measurement of these side-bands can be used to infer the particle's current position after filtering out the elastically scattered signal. This is a non-intrusive set-up that could be implemented in magnetic traps to give a significant increase in measurement efficiency and resulting position resolution, over current imaging schemes.

The interaction Hamiltonian, considering only emission into the mirror mode, is
\begin{equation}
H_{\text{I}}=\hbar\sqrt{\gamma}\sin(k_{L}(L+\hat{x}))\big(b+b^{\dagger}\big)\text{.}
\end{equation}
If the position of the trap centre is taken to be where $k_{L}L=\pi/4$, we can define the corresponding system operator
\begin{equation}
\hat{c}=\sin\big(k_{L}(L+\hat{x})\big)\approx\frac{1}{\sqrt{2}}\big(1+k_{L}\hat{x})\big)\text{,}
\end{equation}
\noindent
where we have performed a Taylor expansion in the Lamb-Dicke regime. This expansion is possible when the typical length of the oscillation is small compared with the wavelength of incident light, $k_{L} x \ll 1$ (some initial cooling would be required to reach this regime). We note that this operator has two separate components, describing the effects of constant amplitude elastically scattered light and position-dependent modulated light. 

We can then apply continuous measurement theory from quantum optics\,\cite{gardiner2004quantum,carmichael2009open} to the system. Under the usual Born and Markov approximations, for this form of the interaction Hamiltonian, we can think of the operator $\hat{c}$ as being applied to the system whenever a photon is emitted into the field $\rho \to \hat{c}\rho\hat{c}^{\dagger}/\langle \hat{c}^{\dagger}\hat{c}\rangle$.
A stochastic increment $dN$ can be used to model whether or not a photon is detected in a given time-step in the environment, taking a value of zero or one respectively. Its average value should be the detection rate,
\begin{equation}
\langle dN\rangle = \gamma \langle \hat{c}^{\dagger}\hat{c}\rangle dt \text{,}
\end{equation}
corresponding to the expected value of measuring a scattered photon in the mirror mode in a time interval $dt$. In the limit where the component of elastically scattered light is comparatively large it is helpful to make a diffusion approximation, as is commonly done when considering homodyne detection\,\cite{Wiseman1993}. This is indeed the case here and so
\begin{equation}
\begin{split}
dN &= \gamma \hat{c}^{\dagger}\hat{c}\, dt \approx \frac{\gamma}{2}dt +\gamma k_{L}\hat{x}dt \\
&= \frac{\gamma}{2}dt + \gamma k_{L}\langle x\rangle dt + \frac{\sqrt{\gamma}}{2} dW \text{,}
\end{split}
\end{equation}
where in the last line we have followed the usual analysis for random events occurring quickly enough to be treated as continuous noise, splitting the increment on the right hand side into a sum of two parts. one  deterministic and the other stochastic. The Wiener increment $dW$ represents Gaussian white noise. This signal corresponds directly to what would be measured experimentally by a photo detector. The resulting state evolution is described by a master equation conditioned on the Gaussian measurement collapses \cite{Jacobs2006}
\begin{equation}
d\rho=-\frac{i}{\hbar}\big[H_{\text{sys}},\rho\big]dt + 2\kappa\,\mathcal{D}[x]\rho \, dt + \sqrt{2\eta k}\,\mathcal{H}[x] \rho \, dW \text{.}
\label{master}
\end{equation}
Here $\mathcal{D}[x]$ is the usual Lindblad super-operator that describes dissipation and $\mathcal{H}[x]$ is the measurement super-operator that localises the particle based on the information gathered;
\begin{equation}
\mathcal{D}[\hat{c}]\rho=\hat{c}\rho \hat{c}^{\dagger}-\frac{1}{2}(\hat{c}^{\dagger}\hat{c}\rho+\rho \hat{c}^{\dagger}\hat{c}) \text{,}
\end{equation}
\begin{equation}
\mathcal{H}[\hat{c}]\rho=\hat{c}\rho+\rho \hat{c}^{\dagger}-\langle \hat{c}+\hat{c}^{\dagger}\rangle \rho \text{.}
\end{equation}
The \textit{measurement strength} $\kappa$ is defined as the ratio between the scattering rate and the reduction in position uncertainty of the particle $\delta\alpha$ due to each photon,
\begin{equation}
\kappa = \frac{\gamma}{\delta\alpha^{2}} = \frac{\gamma k_{L}^{2}}{2} \text{.}
\end{equation}
\noindent
This measurement strength reflects the rate of information gained about the system and the corresponding disturbance this necessarily causes. This exact expression for $\kappa$ would be accurate if the scattering was exclusively along the $x$ axis, the true value will be less in any other case where we should only count the momentum kicks projected along the $x$ direction. This is a small correction and should not be a problem given that $\kappa$ otherwise scales with increasing scattering rate off the particle, and can be adjusted by increasing the laser power. The parameter $\eta$ is the quantum efficiency, and accounts for the fraction of photons collected (after projection along the measurement axis) and any further loss that occurs in the detector. The measured photo-current can be expressed as a renormalisation of the now continuous photon count $\langle dN\rangle$, after subtracting the elastically scattered signal in post-processing,
\begin{equation}
dI = \langle x\rangle dt + \frac{1}{\sqrt{8\eta \kappa}}dW \text{.}
\label{measurement}
\end{equation}

It has been suggested that light collection efficiency of $\eta \approx 0.15$ could be reasonably expected when monitoring an optically trapped ion in front of a mirror\,\cite{Bushev2006}. One of the significant advantages of magnetic levitation is that the illumination light is independent of the trapping mechanism, which allows it to be shaped to optimize detection efficiency. This is of crucial importance when relying on active feedback cooling in order to counteract the random motion induced by the measurement itself. The shot-noise in optically trapped nano-particle experiments currently poses a major obstacle to reaching the ground state, with typical collection efficiencies $\eta < 0.01$\,\cite{Jain2016}.

\section{\label{sec:m&e}Measurement \& State estimation}
\subsection{\label{sec:measurement}Measurement}

The main obstacles to ground state cooling using active feedback are environmental heating mechanisms and the fundamental disturbance associated with making measurements. In order to reach the quantum regime it will be necessary for environmental heating to be made negligibly small on the time-scales of the measurement and feedback. A reasonable goal would be to cool a particle in a time comparable to the oscillation period of a $\omega=2\pi\times100$Hz trap. In this case the phonon reheating rate would need to be reduced to around $\Gamma_{\text{th}}=k_{B}T\gamma_{\text{th}}/\hbar\omega\sim1$Hz, where $T$ represents the surrounding gas temperature and $\gamma_{\text{th}}$ is the thermal damping rate. 
Current typical reheating values are around $100$Hz, and below 10 mbar, thermal decoherence is expected to be linear in gas pressure and in the temperature of the environment. By better isolating the particle, or with the help of cryogenically cooling the trap chamber, reheating rates two orders of magnitude lower could feasibly be reached. Attempting to cool on shorter time-scales comes with its own physical limitations, and requires stronger measurements which are in turn a separate source of heating.


It is helpful to consider the necessary measurement strength to reach a desired position resolution in a given time. A simple estimate of the resolution achievable across an interval $\Delta t$, can be found be integrating the measurement record \cite{Doherty2012},
\begin{equation}
\Delta I = \int_{t}^{t+\Delta t} dI \approx \langle x\rangle \Delta t + \int_{t}^{t+\Delta t} \frac{dW}{\sqrt{8\eta \kappa}} \text{.}
\end{equation}
In this expression we have assumed that the expected value of the position of the particle will not change much over the time interval. This is not a well justified assumption but will allow us to determine an upper bound for the resolution. The integrated measurement signal $\Delta I$ has a mean value of $\sqrt{8\eta \kappa}\langle x\rangle \Delta t$ that grows linearly in time, and its width grows as the square root, $\sigma=\sqrt{\Delta t}$. Continuous measurement over this interval could therefore resolve at best,
\begin{equation}
\delta x \approx \frac{1}{\sqrt{8 \Delta t\, \eta \kappa}} \text{,}
\label{resolution}
\end{equation}
with a signal to noise ratio of one. We would like to achieve resolution comparable to the size of the quantum ground state $x_{0} = \sqrt{\hbar/2m\omega}$, in some time interval which for now we will consider to be on the order of a mechanical oscillation $\Delta t = 1/\omega$, to outpace a realistic thermal heating rate,
\begin{equation}
\delta x_{\omega} = \sqrt{\frac{\omega}{8\eta \kappa}} \equiv x_{0} \text{.}
\end{equation}
From this, we can conclude in order to approach ground state cooling, it is necessary for $\kappa/x_{0}^{2} \sim \omega/8\eta$. This places a lower bound on the necessary measurement strength, with the trade-off for going to higher values being greater back-action heating and stochastic drift. Actively counteracting the disturbance caused by a probe light, relies on efficiently gathering as much useful information as possible from every scattered photon. This along with the necessary resolution requirement, are the criteria for a suitable measurement. 

We can now assess the merits and shortcomings of various measurement techniques. Camera-like imaging has been used in previous experiments with particles in low frequency traps. A camera follows a particle's position in a plane perpendicular to the direction of light being scattered from it. However, it is light scattered parallel to this plane that imparts the most recoil to the visible motion of the particle. This translates to a very low quantum efficiency. For example, $15\%$ light collection efficiency from a radiating point dipole $f(\theta) = 3/4 \cos(\theta)$, translates to detecting $\sim 1\%$ of the imparted recoil in the imaging plane. Meanwhile a measurement of a particle's motion parallel to the light being scattered, with the same collection efficiency, translates to detecting around $\sim 19\%$ of the relevant recoil (as in Fig.~\ref{fig:setup}). Even so, imaging is simple to implement and for the purpose of initially damping the position variance to around a fraction of a micron, low quantum efficiency will not be an issue. For comparison, a $0.1\mu m$ diameter diamond in a trap $\omega = 2\pi\times100$Hz, will only be quantum limited when approaching the ground state variance of roughly $x_{0}\approx 0.1nm$. Many high efficiency measurements capable of resolving beyond optical-wavelength amplitude motion, require the particle to already be tightly confined. In a large trap this necessitates some initial cooling so that the particle does not move outside the range of these measurement techniques. Feedback using imaging measurements is well suited for this.

Introducing a cavity around the suspended particle is often a reliable way to improve light collection efficiency. Homodyning light from a standing wave cavity can be used to efficiently track the position of a particle, however this necessarily introduces a dipole potential tied to the measurement strength, and has its own associated challenges\,\cite{Steck2006}. Sideband cooling with near resonance light within a cavity has also been proposed as a useful aid in achieving ground state cooling\,\cite{Genoni2016}. However, this would not be compatible with the efficient on-axis light collection available in magnetic traps, and under optimal conditions, stops being beneficial for cooling compared to active feedback alone when $\eta\sim0.2$. This level of efficiency would hopefully be surpassed in future experiments with enhanced directional scattering. A sensitive velocity measurement was proposed for ion cooling by exploiting electromagnetically induced transparency\,\cite{Rabl2005}. This phenomenon could be observed in a travelling wave cavity with a diamond containing an NV centre, however, the velocity information would only be contained in the spontaneously emitted radiation from a necessarily weakly excited state. For a very massive particle this would be an extremely weak measurement $\kappa/x_{0}^{2}\ll \Gamma_{\text{th}}$, unable to suitably resolve the particle for damping on short time-scales. As discussed in the model section, the most suitable method we have found, involves measuring the amplitude modulation of a standing wave due to a particle's motion in front of a single mirror. This technique has been successfully demonstrated with trapped ions\,\cite{Steixner2005,Bushev2006} and has the potential to be very effective for monitoring magnetically levitated nanoscopic particles, when combined with initial cooling of the oscillation amplitude to around a single optical wavelength.

\subsection{\label{sec:estimation}State estimation}

\begin{figure}[ht]
\includegraphics[scale=0.7,trim={0 0 0 1.3cm},clip]{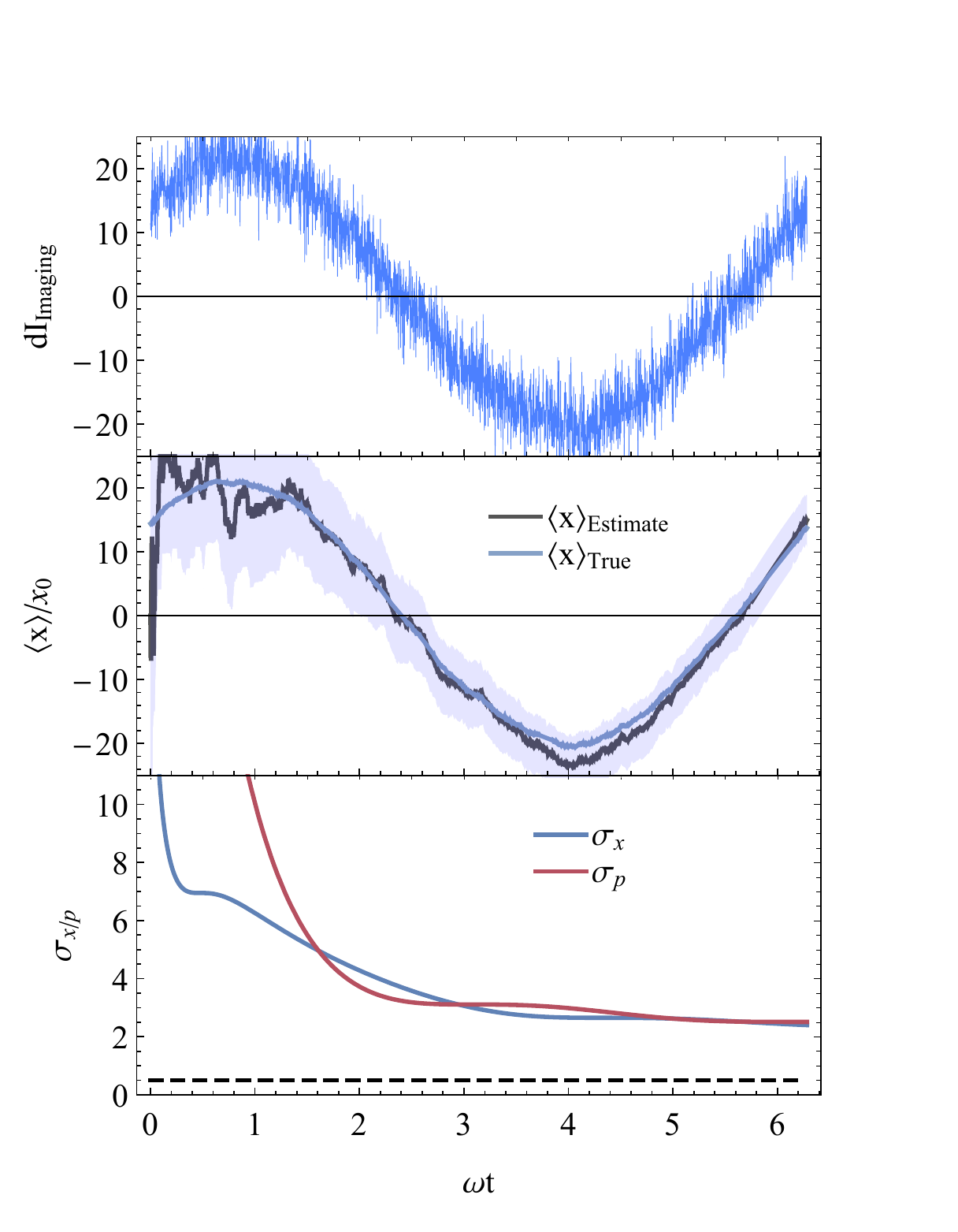}
\caption{Simulation of a trapped particle undergoing measurement, using (\ref{x}-\ref{Cxp}). The normalised measurement strength, $\kappa x_{0}^{2} / \omega = 1$, with $0.1\%$ quantum efficiency, and an initial particle energy corresponding to a temperature of $T=1\mu K$. The top figure shows a numerically generated example of a position measurement and the middle figure shows the results of continuous state estimation using the same signal. The estimated mean position plotted beside the true value, and the shaded region covers 2 standard deviations in the estimate. The bottom figure shows the improvement in the standard deviation in both position and momentum due to the measurement, and the dashed line here indicates the width of the motional ground state.}\label{fig:measurement}
\end{figure}

\begin{figure}[ht]
\includegraphics[scale=0.7,trim={0 0 0 1.3cm},clip]{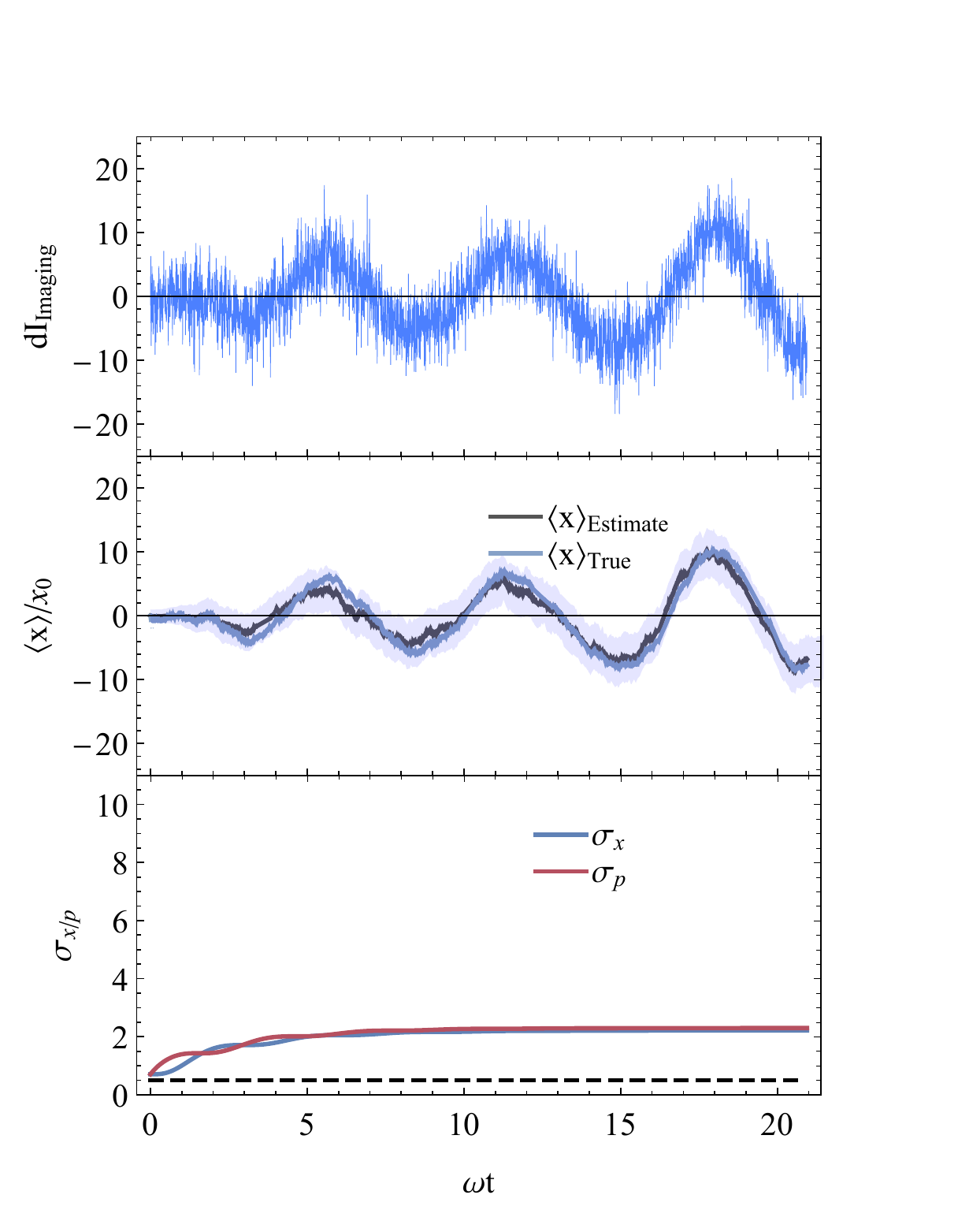}
\caption{Simulation of particle heating due to measurement over several oscillation cycles, using (\ref{x}-\ref{Cxp}). The normalised measurement strength, $\kappa x_{0}^{2} / \omega = 1$, with $0.1\%$ quantum efficiency.
The particle is initially in its ground state with temperature $T=0K$. This figure is otherwise organised in the same way as Fig. \ref{fig:measurement}.}
\label{fig:heating}
\end{figure}

It will be necessary to process the measurement signal in order to perform feedback cooling, since it is not possible to achieve damping by making shifts in the system Hamiltonian proportional to the position alone. Using the equations of motion that describe the particle, combined with the measurement record, the full system state can be continuously estimated. This type of information processing can quickly converge on both the true position and momentum values of the particle, whilst updating the expected error in the estimation.

Using the master equation (\ref{master}), and the fact that $d\langle c\rangle = Tr[c\, d\rho]$, we can find equations of motion for the relevant position and momentum moments to describe a Gaussian state undergoing measurement,

\begin{equation}
d\langle x\rangle = \frac{1}{m} \langle p\rangle dt + \sqrt{8\eta \kappa}\,V_{x}\,dW \text{,}
\label{x}
\end{equation}
\begin{equation}
d\langle p\rangle = -m\omega^{2}\langle x\rangle dt + \sqrt{8\eta \kappa}\,C_{xp}\,dW \text{,}
\label{p}
\end{equation}
\begin{equation}
\partial_{t}V_{x} = \frac{2}{m} C_{xp} - 8\eta \kappa\,V_{x}^{2} \text{,}
\label{Vx}
\end{equation}
\begin{equation}
\partial_{t}V_{p} = -2m\omega^{2}C_{xp} + 2\hbar^{2}\,\kappa - 8\eta \kappa\,C_{xp}^{2} \text{,}
\label{Vp}
\end{equation}
\begin{equation}
\partial_{t}C_{xp} = \frac{1}{m}\,V_{p} - m\omega^{2}V_{x} - 8\eta \kappa\,V_{x}\, C_{xp} \text{.}
\label{Cxp}
\end{equation}

where $V_{x}$ and $V_{p}$ are the position and momentum variances, and $C_{xp} = (1/2)\langle[x,p]_{+}\rangle -\langle x\rangle\langle p\rangle$ is symmetrised covariance. The stochastic increments here can be re-written in terms of the measurement record $dI$, and the equations can be solved to estimate the particle's full motional state. This process would need to be carried out on time-scales $\delta t \ll 1$ms in a $\omega = 2\pi\times 100$Hz trap. The particle's motion is expected to look thermal when cooling starts and this provides a good guess for the particle's initial state variances. The measurement process itself also drives any state towards looking Gaussian, ensuring the continued reliability of these state equations.
This procedure is not dissimilar to estimating the velocity by taking the derivative of the position signal, by passing it through a suitable band pass filter. In fact these state equations are exactly equivalent to the Kalman equations for a noisy classical system, and do indeed act like filters but with dynamic quality factors and cut off frequencies. Kalman equations are designed to update information about a system based on a series of imperfect measurements, and produce an estimate of the system that improves with time better than a series of measurements being made independently\,\cite{jacobs1993introduction}. 

The effectiveness of estimating the state of a levitated particle over a single oscillation cycle, is illustrated in Fig.~\ref{fig:measurement}, for a general position measurement. The true state is numerically modelled using the Gaussian moment equations (\ref{x} - \ref{Cxp}), with an initial temperature of $1\mu K$, which might be realistically achieved with classical feedback damping. The stochastic measurement record (\ref{measurement}) is also numerically generated based on the the current true state. This is then used to update a second set of the same Gaussian moment equations to simulate the state estimation procedure. The state estimate is initiated with thermal variances, where as the true state is modelled as a coherent state with thermal energy. The estimator quickly converges on the true state of the system, until reaching the resolution limit set bu the measurement strength and quantum efficiency. This full state model confirms the rough resolution limit (\ref{resolution}).

The Gaussian state equations can also be used to illustrate the heating effects due to the measurement itself Fig.~\ref{fig:heating}. For the considered measurement strengths this is more easily visible with a state initially prepared at $T=0K$. Without any other sources of environmental heating, the measurement will add energy into the system until it reaches a temperature associated with the magnitude of the photon shot noise. This temperature is higher with more intense illumination and presents a trade off when trying to achieving a better resolution.

\section{\label{sec:analysis}Feedback Cooling}

There are two well established approaches to applying feedback that take into account the effects of quantum noise; direct feedback of a force proportional to the measurement signal\,\cite{Wiseman1994}, and feedback based on real time state estimation\,\cite{A.C.Doherty1999}. It is important to know whether feedback should be treated as direct in order to correctly account for how the noise in the measurement and in the system will be correlated. The simplest approach to damping is to apply a force proportional and opposite to a particle's current velocity, and if measuring the velocity explicitly, this can be implemented as direct feedback\,\cite{Rabl2005}. Similarly, in the case of a high quality oscillator it is sufficient to feedback a signal proportional to the slowly varying momentum quadrature\,\cite{Doherty2012}. Both these techniques require cooling over at least hundreds of oscillation cycles, which is not feasible in low frequency traps. In this case, indirect feedback using the state estimation is necessary, where the low trap frequencies will in fact be beneficial.

\subsection{\label{sec:feedback}Feedback procedure}

The optimal feedback strategy can be determined using classical control theory. In a classical system there would not be noise fundamentally linked to the measurement strength, but this can be artificially enforced. This is useful because it allows well developed control methods to be adapted for cooling\,\cite{Steck2006,Steck2004,Doherty1998}. Our sketch of the idea follows closely the work in Ref.~\cite{A.C.Doherty1999}. 

For this system it turns out not to be optimal to include the estimated state variances in the feedback function. They will be necessary to continuously solve for the mean position and momentum but the feedback will not directly involve the variance values. The feedback Hamiltonian should simply be some linear function of the momentum and position operators scaled by functions of the estimated first order moments,
\begin{equation}
H_{f} = f(\langle x\rangle,\langle p\rangle) x + g(\langle x\rangle,\langle p\rangle) p \text{.}
\end{equation}
To find the appropriate form of the functions $f$ and $g$ we can define a cost function for the parameter we want to minimise, in this case the energy,
\begin{equation}
C = \int_{0}^{t}\big[Tr(\mathbf{x}^{T}P\mathbf{x}\rho) + q^{2}\mathbf{u}^{T}Q\mathbf{u}\big] \text{.}
\end{equation}
Here $\textbf{x} = \{x,p\}$ is the state vector, and $\textbf{u} = -K\langle\textbf{x}\rangle$ is the feedback vector we want to introduce in the dynamical equations for the mean moments (\ref{x},\ref{p}); the optimal form of the matrix $K$ is what needs to be determined. The matrices $P$ and $Q$ are chosen so that the cost function represents the system energy,
\begin{equation}
P=Q=
\begin{pmatrix}
	m\omega^{2} & 0\\ 
	0 & 1/m 
\end{pmatrix}
\text{.}
\end{equation}
The matrix $Q$ can be interpreted as accounting for an energy cost associated with the feedback. Including it in this way reflects a restriction on the magnitude of the feedback weighted by the parameter $q$, which will work out to be inversely proportional to the system damping rate.

Optimal feedback should attempt to localise both position and momentum simultaneously. This is not often a viable option due to the difficulty in creating terms proportional to the momentum operator in the Hamiltonian. A position term in the Hamiltonian can be introduced simply by using an externally applied force. One option to introduce a momentum operator, would be to shift the origin of the position coordinates, which in the rest frame of the trap manifests itself as a shift to the canonical momentum;
\begin{equation}
p \to m(\dot{x} + v) \text{,}
\end{equation}
\begin{equation}
H_{\text{sys}} \to \frac{p^{2}}{2m} -pv +\frac{m \omega^{2} x^{2}}{2} \text{.}
\end{equation}
Where $v$ is the velocity at which the trap centre is shifted. This could be implemented in a low frequency magnetic trap either mechanically or with extra applied fields. The shifts would have to be small, given the measurement's sensitivity to where the particle sits in the standing wave field, but a piezoelectric device could be used to shake the trap in a controlled manner to achieve damping. This would be a unique level of control over both position and momentum for a nano-mechanical system.

Assuming that this could be successfully implemented, the optimal feedback Hamiltonian takes the form,
\begin{equation}
H_{f} = \frac{1}{q} \big(\langle p\rangle x + \langle x\rangle p\big) \text{.}
\end{equation}
We can define $\Gamma = 1/q$, to be the system damping rate, and the parameter $q$ can now be interpreted as a bound on the feedback response time. This accounts for the physical limitations of the feedback mechanism, and places an upper bound on the optimal damping rate. For an infinitely broadband signal $q\to 0$ and the damping rate could be arbitrarily high. With feedback, the new equations for the damped position and momentum are
\begin{equation}
d\langle x\rangle = \frac{1}{m} \langle p\rangle dt + \sqrt{8\eta \kappa}\,V_{x}\,dW - \Gamma \langle x\rangle \text{,}
\label{dx}
\end{equation}
\begin{equation}
d\langle p\rangle = -m\omega^{2}\langle x\rangle dt + \sqrt{8\eta \kappa}\,C_{xp}\,dW - \Gamma \langle p \rangle\text{.}
\label{dp}
\end{equation}

\subsection{\label{sec:results}Cooling results}

\begin{figure}[ht]
	\begin{center}
		\textbf{}\par\medskip
		\includegraphics[scale=1]{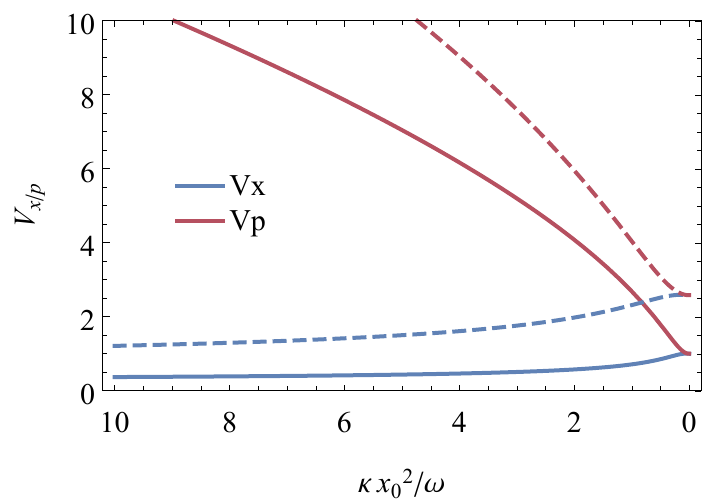}
		\caption{Final resolution of the normalised position and momentum variances of trapped particle, from the steady state solutions of a Guassian estimator (\ref{fVx},\ref{fVp}). Variance values less than 1 are squeezed compared to the harmonic oscillator ground state. The solid lines correspond to a measurement with perfect efficiency $\eta=1$ and the dashed lines $\eta=0.15$, these values and the measurement strength would vary depending on the nature of the measurement.}
		\label{fig:var}
	\end{center}
\end{figure}

In this system the introduction of linear feedback has no effect on the estimated variances conditioned on the measurement record. Their dynamics are governed by the measurement alone and we can therefore find the steady state values for our feedback controlled state from the original equations for the Gaussian moments (\ref{Vx},\ref{Vp},\ref{Cxp}),
\begin{equation}
\tilde{V}_{x} = \frac{2m\omega}{\hbar}\, V_{x} = \bigg(\frac{2}{\eta}\,\frac{1}{\xi + 1}\bigg)^{1/2} \text{,}
\label{fVx}
\end{equation}
\begin{equation}
\tilde{V}_{p} = \frac{2}{\hbar m\omega} \,V_{p} =\bigg(\frac{2}{\eta}\,\frac{\xi^{2}}{\xi + 1}\bigg)^{1/2} \text{,}
\label{fVp}
\end{equation}
where $\xi = \sqrt{1+4/\eta \chi^{2}}$ and $\chi=m \omega^{2}/2\hbar\eta \kappa$. These normalised variances are equal to one for a minimum uncertainty state. This is the case for unit efficiency and when the parameter $\xi \to 1$, which in turn is the case when the measurement strength $\kappa\to0$. Relative to the trap frequency in optical traps, $k$ is usually very small, but with a strong measurement $\kappa x_{0}^{2} > \omega$, the steady state position variance is noticeably squeezed compared to the harmonic oscillator's natural ground state. Fig\,\ref{fig:var} shows how the conditional variances vary for the range of measurement strengths accessible in low frequency magnetic traps.

\begin{figure}[t]
	\begin{center}
	
	\includegraphics[scale=0.7,trim={0 0 0 1.3cm},clip]{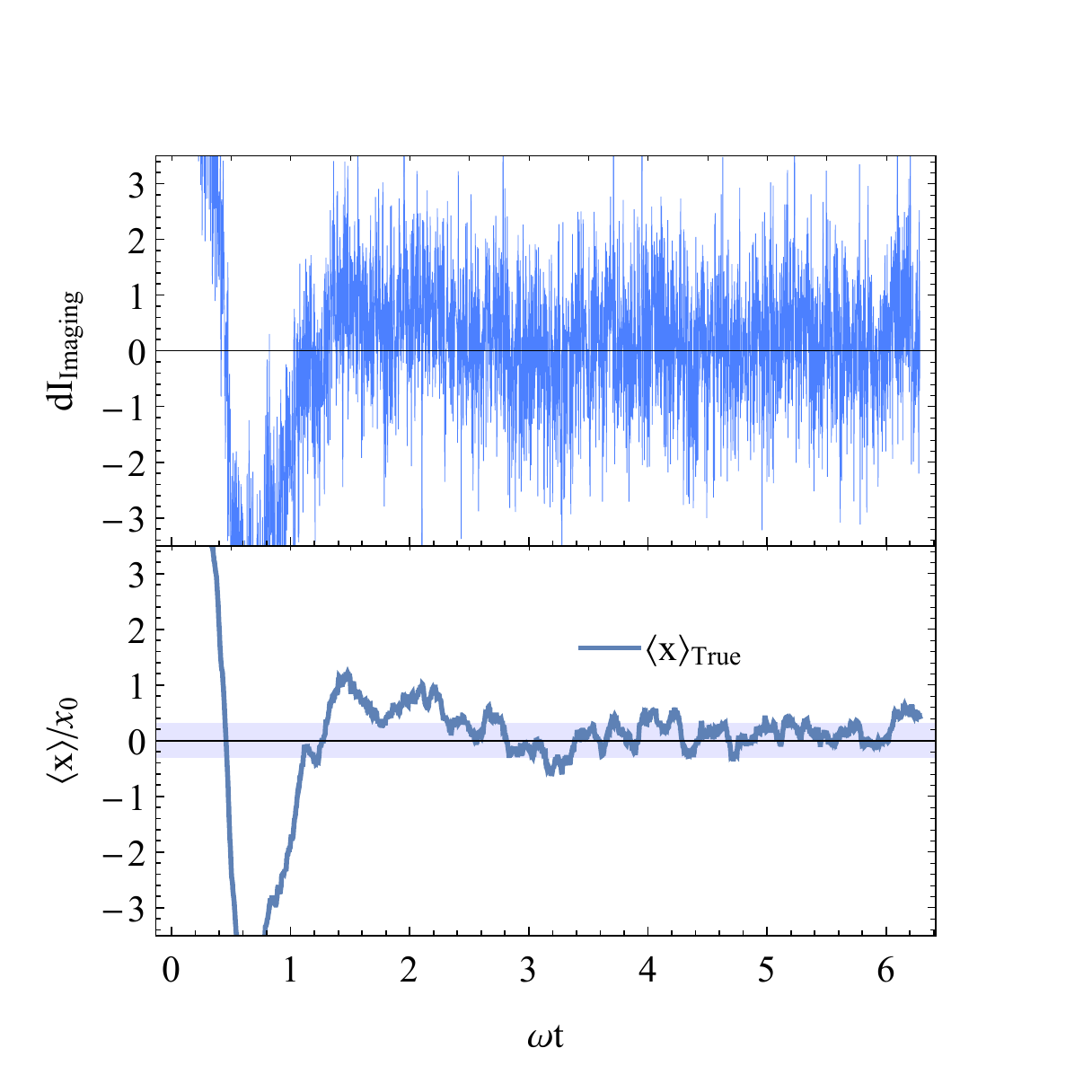}
		\caption{Simulation of a damped levitated particle, using (\ref{dx},\ref{dp},\ref{Vx},\ref{Vp},\ref{Cxp}). The normalised measurement strength, $\kappa x_{0}^{2} / \omega = 1$, with $10\%$ quantum efficiency, and initial particle energy corresponding to a temperature of $T=1\mu K$. The top figure shows a numerically generated example of a position measurement. The bottom figure shows the evolution of the mean position of the true state. The standard deviation of the motion from $t=\pi \to 2\pi$ is highlighted and can be seen to be significantly smaller than the fundamental shot noise in the original measurement signal.}
		\label{fig:damping}
	\end{center}
\end{figure}

\begin{figure}[t]
	\begin{center}
		\textbf{}\par\medskip
		\includegraphics[scale=0.75]{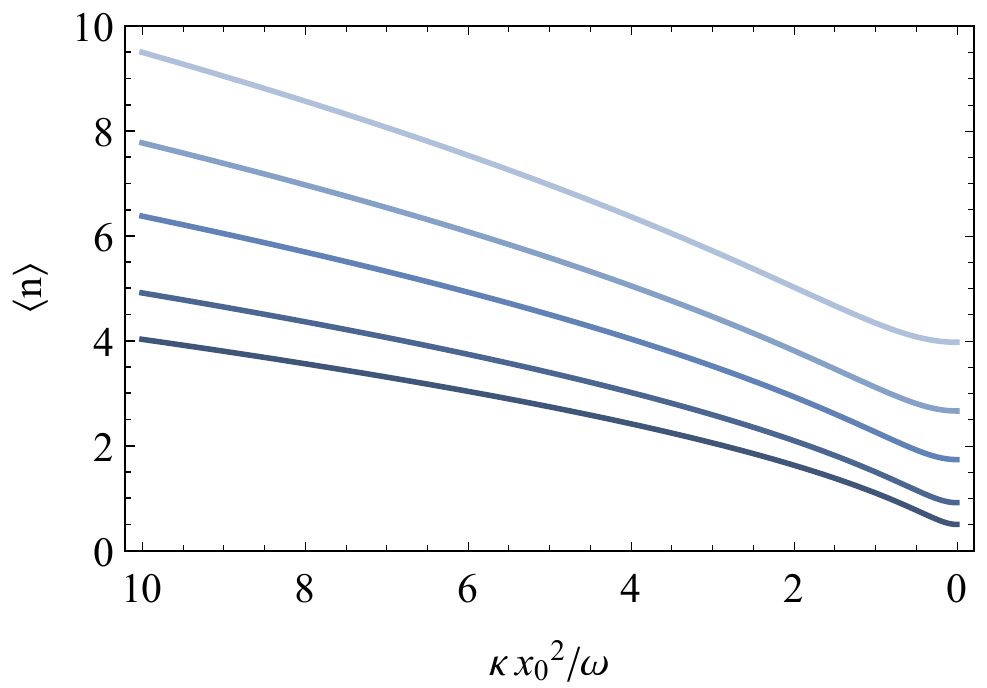}
		\caption{Average steady state phonon occupancy of a trapped nano-particle after undergoing active feedback, calculated using the equations for a damped Gaussian state with excess noise (\ref{phonon}). The effective damping rate (feedback gain) was chosen to be $\Gamma/\omega = 10$, strong enough to remove almost all stochastic drift due to the measurement disturbance. The quantum efficiencies from the top line down are, $\eta = 0.05, 0.1, 0.2, 0.5, 1$. The final occupancies range from $\langle n\rangle <3$, for currently feasible experimental parameters ($\eta=0.2,k=1$), to near zero with perfect collection efficiency and a weaker measurement.}
		\label{fig:limits}
	\end{center}
\end{figure}

\noindent
The estimated variances are the best that could be resolved with a given measurement. We can then average over the measurement record to account for the excess variance due to the particle's motion. The applied feedback should limit this as much as possible, keeping the mean position and momentum values centred on zero. Using the equations for the mean position and momentum (\ref{dx},\ref{dp}), and following the rules of Ito calculus, we can calculate the excess variances, which we have distinguished with a superscript `$E$',
\begin{equation}
\partial_{t}\tilde{V}_{x}^{E} = -2\Gamma \tilde{V}_{x}^{E} + 2\omega\tilde{C}_{xp}^{E} + \frac{2\omega}{\chi}\tilde{V}_{x}^{2} \text{,}
\label{VxE}
\end{equation}
\begin{equation}
\partial_{t}\tilde{V}_{p}^{E} = -2\Gamma \tilde{V}_{p}^{E} - 2\omega\tilde{C}_{xp}^{E} + \frac{2\omega}{\chi}\tilde{C}_{xp}^{2} \text{,}
\label{VpE}
\end{equation}
\begin{equation}
\partial_{t}\tilde{C}_{xp}^{E} = -2\Gamma \tilde{C}_{xp}^{E} - \omega\big(\tilde{V}_{x}^{E}-\tilde{V}_{p}^{E}\big) + \frac{2\omega}{\chi}\tilde{V}_{x}\tilde{C}_{xp} \text{.}
\label{CxpE}
\end{equation}
The final state is always improved with stronger damping which effectively counteracts the measurement shot noise, as well as removing the initial thermal energy. The return for increasing $\Gamma$ quickly drops off, and for moderate damping rates $\Gamma > \omega$ the steady state variances approach the ideal limits given by the measurement resolution. This is reassuring since physically there would certainly be a bound to the feedback response time. Fig\,\ref{fig:damping} shows a simulation of the feedback procedure for experimentally reasonable parameters $\eta=0.1, T_{\text{initial}}=1\mu K, k\omega/x_{0}^{2} = 1, \Gamma=10$. The state is again modelled as a coherent state with thermal energy and feedback is applied based on a numerically simulated state estimator. The particle's motion is almost completely damped after a single oscialltion cycle and the excess variance in the mean position is highlighted, $\tilde{V}_{x}^{E} \sim 0.1$. The remaining motion is small compared to the fundamental resolution limit due to the photon shot noise.

From the steady state expressions we can also find the purity of the final state\,\cite{Zurek1993},
\begin{equation}
Tr(\rho^{2}) = (\hbar/2)(V_{x}V_{p}-C_{xp}^{2})^{-1/2} \text{.}
\end{equation}
If the damping is strong, the steady state value is approximately that of a conditional state without any excess. With perfect detection the final measured state looks pure, and becomes increasingly mixed as the efficiency drops,
\begin{equation}
P_{c}=Tr(\rho_{c}^{2}) = \sqrt{\eta} \text{.}
\end{equation}
To reach the lowest temperatures, $k$ would ideally be kept as low as possible to avoid squeezing due to the measurement. There is a balance then between resolving the particle fast enough to outpace environmental heating, and wanting a weak probe to minimise squeezing. Notably however, state purity has no dependence on the measurement strength, suggesting that the squeezed states with higher energy could reasonably be expected to have quantum properties which are just as visible.

The final average phonon number can be calculated using the combined conditional variances based on a particular measurement, and the excess variance seen when averaging over trajectories,
\begin{equation}
\langle n\rangle = \frac{\langle x^{2}\rangle}{2}+\frac{\langle p^{2}\rangle}{2}-\frac{1}{2} \text{.}
\label{phonon}
\end{equation}
Steady state phonon occupancy, calculated with (\ref{phonon}), is shown in Fig\,\ref{fig:limits}, for a range of measurement strengths and quantum efficiencies. These are the expected values that would be observed after damping, taking into account the estimated variance in the measurement signal (\ref{fVx},\ref{fVp}), and the excess variance associated with the remaining particle motion (\ref{VxE}-\ref{CxpE}).

\noindent

\section{\label{sec:Conclusions}Conclusions and Outlook}

In this article, we have analysed processes for state estimation and feedback cooling of a low-frequency, magnetically levitated nano-particle. Monitoring the particle's position through modulation of a standing wave in front of a mirror was chosen as the most suitable option, over monitoring the light output from a cavity. This should be relatively simple to integrate into current experiments, and would allow for a high degree of variation in the measurement strength which would be primarily dependent on the intensity of the probe beam. The need to damp both the particle momentum and position independently is likely to be the largest experimental difficulty after achieving sufficient isolation from environmental heating. The unique nature of the static magnets that make up these traps may make it possible to control the particle by dynamically shifting the trap centre, and alternate methods using a sequence of strong controlled laser pulses are also possible.  

We suggest that measurement efficiency comparable to or greater than that achievable in ion traps, $\eta = 0.15$, could realistically be reached in an experiment. Optimal feedback via state estimation with this level of efficiency could produce states competitively near the quantum ground state with some additional degree of squeezing, $\langle n\rangle < 3$, with purity $P\approx0.44$, in only a few oscillation periods. In current experiments there are many factors to consider in order to extend the system reheating time, which will be the main barrier to achieving lower temperatures as it prevents the use of a less disruptive measurement probe. As these values improve, and with the possibility of highly directional scattering for better collection efficiency, it may soon be feasible to reach below single phonon occupancies using the methods outlined in this article. Most related experiments have so far assessed success based on a temperature associated with the measured motional power spectrum. Alternatively, there are recent proposals for distinguishing quantum motion via dynamical model selection using solely position measurements\,\cite{Ralph2018}. They look to identify quantum statistics from a series of position measurements after introducing a small perturbation to the trapping potential. The distinguishability is closely related to state purity, which should be safely within reach of the proposed cooling methods.

All of the methods discussed are applicable to sub-micron sized Rayleigh scatterers that can be effectively treated as point dipoles. High quality nano-diamonds of this size have been produced for exactly the purpose of trapping and cooling\,\cite{Frangeskou2018}. Microscopic particles on the other hand would not usually be suitable for the sub-wavelength measurements suggested. However, large diamonds could still be cooled by tracking the position of point-like NV impurities within them. 

Additionally, strong coupling between an NV spin and the mechanical oscillation of a nano-diamond can be engineered using a strong magnetic field gradient. There are proposals for generating low number Fock states and possible spatial superposition states, by manipulating a Jaynes-Cummings type interaction Hamiltonian, in states prepared near the quantum ground state \cite{PhysRevA.88.033614}.

\section*{Acknowledgements}
The authors thank B. D'Urso, B.R. Slezak, C.W. Lewandowski and P. Nachman for discussions which motivated the present work, and for their helpful comments regarding experimental details. L.S.W. acknowledges support from the EPSRC through a DTP studentship and SUPA for a PECRE award to visit Montana State University.

%

\end{document}